\documentclass[fleqn,twoside]{article}
\usepackage{espcrc2}

\usepackage{graphicx}
\usepackage[figuresright]{rotating}

\newcommand{\AmS}{{\protect\the\textfont2
  A\kern-.1667em\lower.5ex\hbox{M}\kern-.125emS}}

\title{
Radiation measurements in the new tandem accelerator FEL }

\author{
A.Gover\address[TAU]{ Deparment of Physical Electronics, Faculty
of Engineering, Tel Aviv University, Tel Aviv, Israel },
A.Faingersh\addressmark[TAU], A.Eliran\addressmark[TAU],
M.Volshonok\addressmark[TAU], H.Kleinman\addressmark[TAU],
S.Wolowelsky\addressmark[TAU], Y.Yaakover\addressmark[TAU],
B.Kapilevich\address[Jud&Sh]{ Deparment of Electrical and
Electronic Engineering, The College of Judea and Samaria, Ariel,
Israel }, Y.Lasser\addressmark[Jud&Sh],
Z.Seidov\addressmark[Jud&Sh], M.Kanter\addressmark[Jud&Sh],
A.Zinigrad\addressmark[Jud&Sh], M.Einat\addressmark[Jud&Sh],
Yu.Lurie\addressmark[Jud&Sh], A.Abramovich\addressmark[Jud&Sh],
A.Yahalom\addressmark[Jud&Sh], Y.Pinhasi\addressmark[Jud&Sh],
E.Weisman\address[Rafael]{Rafael, Haifa 31021, Israel}, J.Shiloh\addressmark[Rafael]
}

\begin{document}

\begin{abstract}
The Israeli Tandem Electrostatic Accelerator FEL (EA-FEL), which
is based on an electrostatic Van der Graaff accelerator was
relocated to Ariel 3 years ago, and has now returned to operation
under a new configuration. In the present FEL, the millimeter-wave
radiation generated in the resonator is separated from the
electron beam by means of a perforated Talbot effect reflector. A
quasi-optic delivery system transmits the out-coupled power
through a window in the pressurized gas accelerator tank into the
measurement room (in the previous configuration, radiation was
transmitted through the accelerator tubes with 40 dB attenuation).
This makes it possible to transmit useful power out of the
accelerator and into the user laboratories.

After re-configuring the FEL electron gun and the e-beam transport
optics and installing a two stage depressed collector, the e-beam
current was raised to 2~A. This recently enabled us to measure
 both spontaneous and stimulated emissions of radiation in
the newly configured FEL for the first time. The radiation at the
W-band was measured and characterized. The results match the
predictions of our earlier theoretical modeling and calculations.
\vspace{1pc}
\end{abstract}

\maketitle

\section{Introduction}

The Israeli electrostatic accelerator FEL (EA-FEL) is based on a
6~MeV EN-Tandem Van de Graaff accelerator, which was originally
used as an ion accelerator for nuclear physics experiments
\cite{1}.  The scheme employs straight geometry for the electron
beam transport, where the electron gun and the collector are
installed outside of the accelerator region. Lasing was reported
in a previous configuration, where radiation was transmitted
through the accelerator tubes with 40dB attenuation \cite{2,3}.

In the present version of the FEL, which was relocated to Ariel,
the millimeter-wave radiation generated in the resonator is
separated from the electron beam by means of a perforated Talbot
effect reflector \cite{4,5}. A quasi-optic delivery system
transmits the out-coupled power through a window in the
pressurized gas accelerator tank. The basic parameters of the FEL
are summarized in Table~1. The acceleration voltage is set to be
$E_k=1.4~MeV$ in order to tune the frequency of the FEL radiation
to the W-band near 100~GHz.

In the following sections, we present an analysis and the results
of spontaneous and stimulated emissions measurements carried out
recently.

\begin{table}[tbh]
\caption{Parameters of the tandem electrostatic accelerator FEL}
\label{table} 
\begin{center}
\begin{tabular}
{ll} &\\ \multicolumn{2}{l}{Accelerator} \\
   $\ \ $Electron beam energy: &  $E_k = 1 - 3$~MeV  \\
   $\ \ $Beam current:         &  $I_0 = 1 - 2$~A    \\
&  \\
 \multicolumn{2}{l}{Undulator}        \\
    $\ \ $Type: &  \multicolumn{1}
    {p{0.2\textwidth}}{Magneto-static \newline
    planar wiggler}         \\
   $\ \ $Magnetic induction: &   $B_W$=2~kG        \\
    $\ \ $Period length:      &   $\lambda_W$=4.444~cm  \\
    $\ \ $Number of periods:  &   $N_W$=20              \\
& \\ \multicolumn{2}{l}{Resonator} \\
     $\ \ $Waveguide:&
\multicolumn{1}{p{0.2\textwidth}}{Curved-parallel plates}   \\
 $\ \ $Transverse mode:   &    $TE_{01}$     \\
   $\ \ $Round-trip length:  &    $L_c$=2.62~m  \\
   $\ \ $Out-coupling coefficient:  &  $T$=7\%  \\
   $\ \ $Total round-trip reflectivity: & $R$=65\%  \\
\end{tabular}
\end{center}
\end{table}
\section{Spontaneous emission in a resonator}
Random electron distribution in the e-beam causes fluctuations in
current density, identified as {\em shot noise} in the beam
current. Electrons passing through a magnetic undulator emit a
partially coherent radiation, which is called {\em undulator
synchrotron radiation}. The electromagnetic fields excited by each
electron add incoherently, resulting in a {\em spontaneous
emission} with generated power spectral density \cite{6}:

\begin{equation}
\frac{d P_{sp}(L_W)}{d f} = \tau_{sp} P_{sp}(L_W) \mbox{ sinc} ^2
\left( \frac{1}{2} \theta L_W \right)
    \label{}
\end{equation}
where $P_{sp}(L_W)$ is the {\em expected value} of the spontaneous
emission power, $\tau_{sp} = \left| (L_W/V_{z0}) - (L_W/V_g)
\right|$ is the slippage time and $\theta = (2\pi f/V_{z0}) - (k_Z
+ k_W)$ is the detuning parameter ($V_{z0}$ is the axial velocity
of the accelerated electrons and $V_g$ is the group velocity of
the generated radiation). The spontaneous emission null-to-null
bandwidth is approximately $2/\tau_{sp} = 2 (f_0/N_W)$. In a FEL,
utilizing a magneto-static planar wiggler, the total power of the
spontaneous emission is given by:
\begin{equation}
P_{sp}(L_W) = \frac{1}{8} \frac{e I_0}{\tau_{sp}} \left(
\frac{a_W}{\gamma \beta_{z0}} \right)^2\frac{Z}{A_{em}} L_W ^2
\end{equation}
where $Z \approx 2\pi f \mu _0/k_z$ is the mode impedance, and
$I_0$ is the DC beam current. The expected value of the total
spontaneous emission power generated inside the cavity is about
$P_{sp}(L_W)/I_0 = 60\ \mu W A^{-1}$. The calculated spectrum of
the spontaneous emission power of the Israeli EA-FEL, has a
null-to-null bandwidth of 18~GHz.

At the resonator output, the spontaneous emission spectrum
generated inside the resonator is modified by a Fabry-Perot
spectral transfer-function \cite{7}:

\begin{eqnarray}
\lefteqn{\frac{d P_{out}}{d f} =} &&   \nonumber \\
&& \hspace{-5mm}
 \frac{T}{(1 - \sqrt{R})^2 + 4 \sqrt{R}\sin ^2 \left(k_z L_c/2 \right)}
                              \cdot   \frac{d P_{sp}(L_W)}{d f}
\end{eqnarray}
where $L_c$ is the resonator (round-trip) length, $R$ is the total
power reflectivity of the cavity, $T$ is the power transmission of
the out-coupler and  $k_z(f)$ is the axial wavenumber of the
waveguide mode. The maxima of the resonator transfer function
factor occur when $k_z(f_m) \cdot L_c = 2 m\pi$ (where $m$ is an
integer), which defines the resonant frequencies $f_m$ of the
longitudinal modes. The {\em free-spectral range} (FSR) (the
inter-mode frequency separation) is given by $\mbox{FSR} = v_g/L_c
\approx 113$~MHz. The transmission peak is $T/(1 - \sqrt{R})^2
\approx 1.6$ with {\em full-width half-maximum} (FWHM) bandwidth
of $\mbox{FWHM} = \mbox{FSR/F} \approx 7.76$~MHz, where $F = \pi
\sqrt[4]{R}/(1 - \sqrt{R})=14.56$ is the {\em Finesse} of the
resonator. The spectral line-shape of the spontaneous emission
power obtained at the resonator output of the EA-FEL is shown in
Fig.~1.

The {\em noise equivalent bandwidth} is defined as the bandwidth
of an ideal band-pass filter producing the same noise power at its
output. The noise equivalent bandwidth of a single resonant
longitudinal mode is $B = (\pi/2)\mbox{FWHM} \approx 12.2$~MHz.
Consequently, the spontaneous emission power of mode $m$ is given
by:
\begin{equation}
P^{out}_{sp}(m) =
 \frac{T}{(1 - \sqrt{R})^2} \cdot  \left. \frac{d P_{sp}(L_W)}{d f}
 \right|_{f_m} \cdot B.
\end{equation}
The typical bandwidth of the generated spontaneous emission power
spectrum (1) is $1/\tau_{sp} \approx 9$~GHz. The number of
longitudinal modes within the spontaneous emission bandwidth is
then $N_{modes} = (1/\tau_{sp})(1/\mbox{FSR} \approx 80$. Thus the
total spontaneous emission power measured at the output of the
resonator is given as follows:
\begin{eqnarray}
P^{out}_{sp} &=& N_{modes} \cdot P^{out}_{sp}(m)
                    \nonumber   \\
&\approx&
 \frac{T}{(1 - \sqrt{R})^2} \cdot  P_{sp}(L_W)
\end{eqnarray}

Using equation (2), we expect up to $P^{out}_{sp}(L_w) \approx
120\ \mu\mbox{W}$ spontaneous emission power to be generated
inside the resonator. From (5), the power emitted from the
resonator out-coupler is reduced to $P^{out}_{sp} \approx 24\
\mu\mbox{W}$. The attenuation of the wave-guiding system,
delivering the power from the resonator, located inside the
high-voltage terminal, to the measurement apparatus is 10dB.
Consequently, the spontaneous emission power expected at the
detector sight is 2.4~$\mu$W. The traces shown in Fig.~2, describe
the electron beam current pulse and the signal obtained at the
detector video output, corresponding to the measured spontaneous
emission RF power.

\begin{figure}[tbh]
\centerline{\includegraphics[width=0.5\textwidth,angle=0]{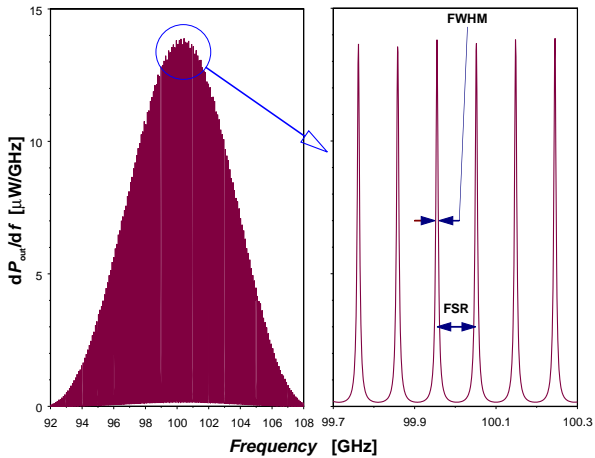}}
\caption{Spontaneous emission power spectrum at resonator output
(for $I_0=1$~A).} \label{Fig1}
\end{figure}

\begin{figure}[tbh]
\centerline{\includegraphics[width=0.5\textwidth,angle=0]{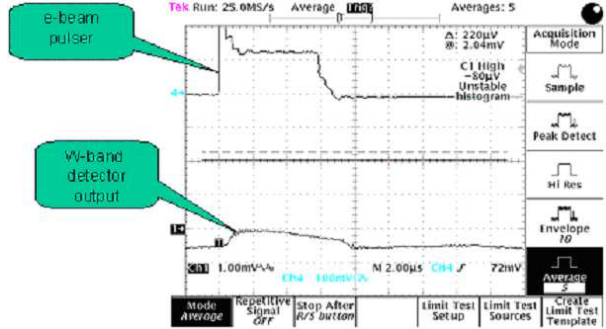}}
\caption{Spontaneous emission power measurement.} \label{Fig2}
\end{figure}

\section{Stimulated emission}

In the present operation regime of the FEL, the efficiency of
energy extraction from the electron beam is given in terms of the
number of wiggler's periods $N_w$  by the approximate formula
$\eta_{ext} \approx 1/2N_W = 2.5$~\%. The stimulated radiation
power generated inside the resonator at steady state is given as
follows:
\begin{equation}\Delta P = \eta_{ext} E_k I_0 \end{equation}
where $\Delta P \approx 35$~kW for a beam current of $I_0 = 1$~A.
The resulted power obtained from the out-coupler is given as
follows:
\begin{equation}P_{out} = \frac{T}{1 - R} \Delta P \end{equation}
and evaluated to be $P_out = 7$~kW. Considering the attenuation of
the transmission system, 700~W is expected at the detector. Fig.~3
shows recent measurement of 150~W radiation power at the end of
the optical transmission line in the measurement room. We note
that in the present preliminary experiments, only a fraction of
the cathode current was transported through the wiggler, and no
beam circulation (transport up to the collector) was achieved. The
charging of the terminal caused voltage drop of the terminal of
125~kV during the pulse duration. Evidently, the FEL had not yet
reached saturation because the radiation mode built inside the
resonator went out of synchronism with the beam before reaching
saturation.\\

\begin{figure}[tbh]
\centerline{\includegraphics[width=0.5\textwidth,angle=0]{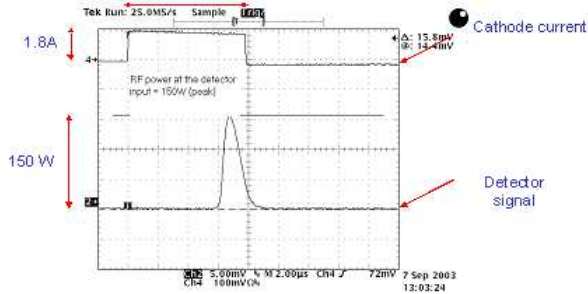}}
\caption{Stimulated emission (lasing) power measurement.}
\label{Fig3}\end{figure}

\noindent {\bf Acknowledgments}\\

This work was carried out at the Israeli FEL National Knowledge
Center supported by The Ministry of Science, Israel, and was
supported in part by the Ministry of Infrastructure.


\end{document}